\documentclass[twocolumn,showpacs,pra,aps]{revtex4}
\usepackage{amsmath}
\usepackage{amsfonts}
\usepackage{amssymb}
\usepackage{graphicx}
\usepackage{hyperref}
\usepackage{bm}
\usepackage{color}

\everymath{\displaystyle} 
\newcommand{\bra}[1]{\ensuremath{\left\langle#1\right\vert}} 
\newcommand{\ket}[1]{\ensuremath{\left\vert#1\right\rangle}} 
\newcommand{\md}[1]{\ensuremath{\left\vert#1\right\vert}} 

\begin{document}

\title{Second order coupling between excited atoms and surface polaritons}
\author{Sofia Ribeiro$^{1}$, Stefan Y. Buhmann$^{1}$ and Stefan Scheel$^{1,2}$}
\affiliation{$^{1}$ Quantum Optics and Laser Science, Blackett Laboratory,
Imperial College London, Prince Consort Road, London SW7 2BW, United Kingdom}
\affiliation{$^{2}$ Institut f\"ur Physik, Universit\"at Rostock, Universit\"atsplatz
3, D-18051 Rostock, Germany}

\date{\today}

\begin{abstract}
Casimir--Polder interactions between an atom and a macroscopic body are
typically regarded as due to the exchange of virtual photons. This is strictly
true only at zero temperature. At finite temperature, real-photon exchange can
provide a significant contribution to the overall dispersion interaction. Here
we describe a new resonant two-photon process between an atom and a planar
interface. We derive a second order effective Hamiltonian to
explain how atoms can couple resonantly to the surface polariton modes of the
dielectric medium. This leads to second-order energy exchanges which we compare
with the standard nonresonant Casimir--Polder energy. 
\end{abstract}


\maketitle

\section{Introduction}

Fluctuation-induced forces such as Casimir--Polder (CP) forces between atoms or
molecules and macroscopic bodies are manifestations of the zero-point energy of
the electromagnetic vacuum \cite{casimirpolderpaper}. They occur even if the
atom and the macroscopic body are in their respective (unpolarized) ground
states \cite{acta2008} and can be understood --- at least in the nonretarded
limit --- as interactions between a spontaneously generated atomic dipole and
its (instantaneous) image inside the macroscopic body. As soon as it became
possible to achieve atom-surface distances below $100\mu$m, experiments revealed
that the coupling between the atom and the surface at these short distances
would produce significant effects \cite{edhindspaper}.
 
Following the advances in laser cooling and trapping techniques in the 1980s, a
new area of research has emerged. Modern laser-based techniques have allowed an
unprecedented amount of control, with this control came the ability to study
very large atom-based systems. As a result of these advances, trapping and
manipulating single atoms, driving atoms into highly excited Rydberg states or
creating Bose-Einstein condensates have become possible. New complex
microstructures like atom chips allow one to trap, cool and manipulate ensembles
of ultracold atoms in the vicinity of a surface \cite{atomchipbook}.

Atoms and surface polaritons are very distinct quantum objects with different
characteristics which make them suitable to perform different tasks. Atoms are
very good candidates for storing and manipulating quantum information. The
extremely promising results in the field of Rydberg atoms, both in ultracold
atoms or in thermal vapours, shown that they make very good candidates to build
quantum gates \cite{NPL5_115_2009,NPL5_110_2009}. The renewed experimental
interest in Rydberg atoms is due to the unique opportunities afforded by their
exaggerated properties \cite{Gallagher} which make them extremely sensitive to
small-scale perturbations of their environment and to dispersion forces.
Previous work \cite{AlexPRA2010} showed that these properties includes massive
level shifts that a Rydberg atom experiences in close proximity of another atom
or in the vicinity of a macroscopic body, with shifts on the order of several
GHz expected at micrometer distances.

Surface polaritons appear at the interface of two media. They represent
particular solutions of the Maxwell equations which correspond to waves
propagating in parallel to the interface and whose amplitude decreases
exponentially when moving away from the surface. They are capable of interacting
and be moved around on a surface, making them very attractive means of
transporting quantum information from one point to another \cite{NP5_494_2011}.
Upon taking advantage of the individual properties of atoms and surface
polaritons and their different properties, it is possible to propose
sophisticated quantum circuits \cite{NP_5_2011}.

Atom-polariton couplings lead to the (nonresonant) Casimir--Polder interaction
between an atom and a planar interface. In the nonretarded limit, this
interaction scales with $1/z^{3}$ ($z$ is the atom-surface distance)
\cite{acta2008}. Moreover, it has already been shown that it is possible to
turn the (usually attractive) Casimir--Polder interaction into a repulsive force
by a resonant coupling between a virtual emission of an atom and a virtual
excitation of a surface polariton \cite{PRL83_26_1999}. Similarly, it has been
shown that the atom-surface coupling can drastically modify atomic branching
ratios, because of surface-induced enhancement of a resonant decay channel
\cite{PRL88_24_2002}.

In this article we analyse a new type of near-field effect involving surface
polaritons inspired by the experiment of K\"ubler \textit{et al.}
\cite{Kubler2010} with hot Rb vapour in glass cells. Their experiment indicated
that a description of the atom-surface interactions should also include a
second order coupling between the atomic transitions and surface polaritons. Their
experimental results indicated that it should be possible for an atom to be
coupled resonantly to the surface polariton modes of the dielectric material
which leads to second-order energy exchanges with the atomic transition energy
matching the difference in polariton energies. 

The article is organised as follows. After briefly reviewing the
formalism of macroscopic QED in Sec.~\ref{sec:basic}, we derive an
effective second order atom-polariton coupling Hamiltonian in
Sec.~\ref{sec:effective} and give concluding remarks in
Sec.~\ref{sec:conclusions}.

\section{Basic Equations}
\label{sec:basic}

In electric dipole approximation, the Hamiltonian that governs the dynamics
of the coupled atom-field system can be written as \cite{acta2008}
\begin{gather}
\hat{H} = \hat{H}_{F} + \hat{H}_{A} + \hat{H}_{\mathrm{int}} \nonumber \\
= \int\limits_0^\infty d \omega \int  d^3r \, \hbar \omega\,
\hat{\mathbf{f}}^\dagger(\mathbf{r}, \omega)\cdot
\hat{\mathbf{f}}(\mathbf{r},\omega) + \sum_n \hbar \omega_n \hat{A}_{nn}
\nonumber \\
- \sum_{m,n} \hat{A}_{nm} \mathbf{d}_{nm} \cdot \hat{\mathbf{E}}(\mathbf{r}_A)
\,.\label{eq:hamiltonian}
\end{gather}
$\hat{H}_{F}$ is the Hamiltonian of the medium--assisted electromagnetic field.
It is expressed in terms of a set of bosonic variables
$\hat{\mathbf{f}}^\dagger(\mathbf{r},\omega)$ and
$\hat{\mathbf{f}}(\mathbf{r},\omega)$ that have the interpretation as amplitude
operators for the elementary excitations of the system composed of the
electromagnetic field and absorbing medium. They obey the commutation rules
\begin{equation}
[\hat{f}_k(\mathbf{r},\omega), \hat{f}^{\dagger}_{k'}(\mathbf{r}', \omega')] =
\delta_{k k'} \delta(\omega-\omega') \delta(\mathbf{r}-\mathbf{r}')\,. 
\end{equation}
$\hat{H}_{A}$ is the free Hamiltonian of an atom with eigenenergies
$E_{n}=\hbar\omega_n$ and eigenstates $\ket{n}$, $\hat{A}_{nm}= \ket{n} \bra{m}$
denotes the transition operators between two internal atomic
states; they obey the commutation rules
\begin{equation}
 [\hat{A}_{kl}, \hat{A}_{mn}] = \delta_{lm} \hat{A}_{kn} - \delta_{kn} \hat{A}_{ml}.
\end{equation}
The most relevant part of the Hamiltonian for our study is the atom--field
interaction Hamiltonian 
\begin{equation}
\hat{H}_\mathrm{int} = - \sum_{n,m} \hat{A}_{nm} \mathbf{d}_{nm} \cdot
\hat{\mathbf{E}}(\mathbf{r}_A),
\end{equation}
with dipole transition matrix elements
$\mathbf{d}_{nm}=\bra{n}\hat{\mathbf{d}}\ket{m}$. The frequency components of
the electric field operator
\begin{equation}
\hat{\mathbf{E}}(\mathbf{r}_{A}) = \int\limits_0^\infty d \omega \,
\hat{\mathbf{E}}(\mathbf{r}_A,\omega) +\mbox{h.c.}
\end{equation}
are constructed via a source-quantity representation from the dynamical
variables $\hat{\mathbf{f}}^\dagger(\mathbf{r},\omega)$ and
$\hat{\mathbf{f}}(\mathbf{r},\omega)$ as
\begin{equation}
\hat{\mathbf{E}}(\mathbf{r}_{A}, \omega) = \int d^3 r \,
\bm{G}_e(\mathbf{r}_{A}, \mathbf{r},\omega) \cdot \hat{\mathbf{f}}(\mathbf{r},
\omega)\,.
\end{equation}
The tensor $\bm{G}_e(\mathbf{r}_{A}, \mathbf{r}, \omega)$ is related to the
classical Green tensor $\bm{G}(\mathbf{r}_{A},\mathbf{r},\omega)$ by
\begin{equation}
\bm{G}_{e}(\mathbf{r}, \mathbf{r}', \omega) = i \frac{\omega^{2}}{c^{2}}
\sqrt{\frac{\hbar}{\varepsilon_{0} \pi} \mathrm{Im}\,
\varepsilon(\mathbf{r}',\omega)} \, \bm{G}(\mathbf{r},\mathbf{r}',\omega),
\end{equation}
where $\varepsilon(\mathbf{r},\omega)$ is the permittivity of the macroscopic
system.
 
The Green tensor itself is a solution of the Helmholtz equation
\begin{equation}
\label{eq:helmholtz}
\left[ \bm{\nabla} \times \bm{\nabla} \times \, -\frac{\omega^{2}}{c^{2}}
\varepsilon(\mathbf{r},\omega) \right] \bm{G}(\mathbf{r},\mathbf{r}', \omega) =
\bm{\delta}(\mathbf{r}-\mathbf{r}')
\end{equation}
together with the boundary condition
$\bm{G}(\mathbf{r},\mathbf{r}',\omega)\to\bm{0}$ for
$|\mathbf{r}-\mathbf{r}'|\to\infty$. The Green tensor obeys the useful integral
relation
\begin{gather}
\int d^3s \,\frac{\omega^{2}}{c^{2}}
\mathrm{Im}\,\varepsilon(\mathbf{s},\omega) \, \bm{G}(\mathbf{r}, \mathbf{s},
\omega) \cdot \bm{G}^\ast(\mathbf{s}, \mathbf{r}', \omega) \nonumber \\
= \mathrm{Im}\,\bm{G}(\mathbf{r},\mathbf{r}', \omega)\,,
\label{eq:magicformula}
\end{gather}
which follows directly from the Helmholtz equation (\ref{eq:helmholtz}) and
which reflects the linear fluctuation-dissipation theorem.

\section{Effective Atom-Polariton Coupling}
\label{sec:effective}

\begin{figure}[ht]
 \includegraphics[width=8cm]{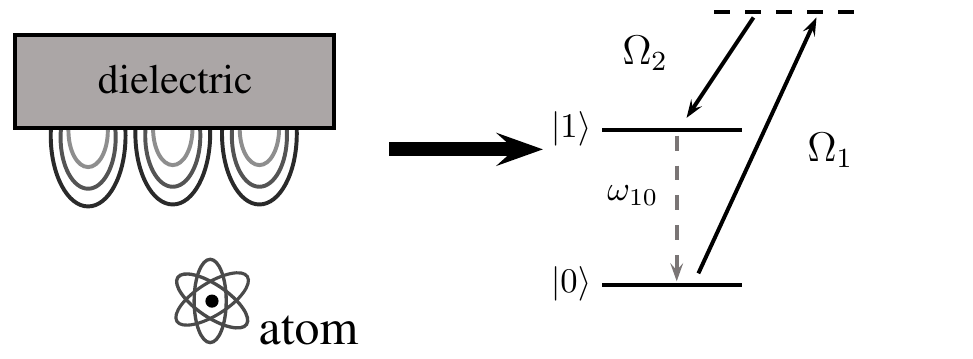}
 \caption{Scheme of resonance between the atomic transition and one surface 
polariton which leads to the creation of a second polariton $\Omega_{1} \approx 
 \omega_{10}+\Omega_{2}$.}
 \label{ourpic}
\end{figure}

In this section, we derive the quantum mechanical description for an effective
second order atom-polariton interaction. The situation we envisage is depicted in
Fig.~\ref{ourpic} in which an atomic transition couples resonantly to two
surface polariton modes of the dielectric material. This corresponds to
second-order energy exchanges with the atomic transition energy matching the
difference in polariton energies.
To illustrate our basic idea, we consider the interaction of an atomic
transition of frequency $\omega_{10}$ between two eigenstates $|0\rangle$ and
$|1\rangle$ with two surface polaritons with corresponding center frequencies
$\Omega_{1}$ and $\Omega_{2}$ $(\Omega_{1} \neq \Omega_{2})$ for whom the
resonance condition $\Omega_{1} \approx \omega_{10}+\Omega_{2}$ is satisfied.
The polariton resonance frequencies $\Omega_{1}$ and $\Omega_{2}$ are assumed to
be far from any other atomic transition frequency $\omega_{mn}$.

Heisenberg's equations of motion
$\dot{\hat{O}}=\frac{i}{\hbar}[\hat{H},\hat{O}]$ for the dynamical variables and
the atomic transition operators follow from the Hamiltonian
(\ref{eq:hamiltonian}) as
\begin{gather}
 \dot{\hat{\mathbf{f}}}(\mathbf{r}, \omega, t) = -i \omega
\hat{\mathbf{f}}(\mathbf{r}, \omega) 
+\frac{i}{\hbar}
\sum_{k,l} \hat{A}_{kl} \mathbf{d}_{kl} \cdot
\bm{G}^\ast_e(\mathbf{r}_{A},\mathbf{r},\omega)
\,, \label{eqf1} \\
\dot{\hat{A}}_{mn}(t) = i \omega_{mn} \hat{A}_{mn} \nonumber \\
-\frac{i}{\hbar} \sum_{k} \left( \hat{A}_{kn} \mathbf{d}_{km} - \hat{A}_{mk}
\mathbf{d}_{nk} \right) \cdot \hat{\mathbf{E}}(\mathbf{r}_{A})\,. \label{eqA1}
\end{gather}
These two equations of motion describe the light atom system, for a full
solution one needs to solve the two equations. Formal integration of
Eq.~(\ref{eqA1}) yields
\begin{gather}
\hat{A}_{mn}(t) = e^{i \omega_{mn} t} \hat{A}_{mn} -\frac{i}{\hbar} \sum_{k}
\int\limits_0^t d t' \: e^{i \omega_{mn}(t-t')}  \nonumber \\
\times \left[ \hat{A}_{kn}(t') \mathbf{d}_{km} - \hat{A}_{mk}(t')
\mathbf{d}_{nk} \right] \cdot \hat{\mathbf{E}}(\mathbf{r}_A,t'),
\label{eqA2}
\end{gather}
which, inserted back into the equation of motion for the dynamical variables of
the medium-assisted field, Eq.~(\ref{eqf1}), yields the first iteration of
the equations of motion for the dynamical variables as
\begin{widetext}
\begin{gather}
\dot{\hat{\mathbf{f}}}(\mathbf{r}, \omega, t) = -i \omega
\hat{\mathbf{f}}(\mathbf{r}, \omega) 
+
\frac{i}{\hbar}
\sum_{m,n} e^{i \omega_{mn} t} \hat{A}_{mn}
\mathbf{d}_{mn}  \cdot \bm{G}^\ast_e(\mathbf{r}_{A}, \mathbf{r},\omega)
+\frac{1}{\hbar^2}
\sum_{k,m,n} \int\limits_0^t d t'
\int\limits_0^\infty d\omega' \int d^3r' \, 
e^{i \omega_{mn} (t-t')} \nonumber \\
\times \mathbf{d}_{mn} \cdot \bm{G}^\ast_e(\mathbf{r}_{A}, \mathbf{r},\omega) 
\left[ \hat{A}_{kn}(t') \mathbf{d}_{km} - \hat{A}_{mk}(t')
\mathbf{d}_{nk} \right] \cdot \left[ \bm{G}_e(\mathbf{r}_{A},
\mathbf{r}',\omega') \cdot \hat{\mathbf{f}}(\mathbf{r}',\omega', t') 
+\bm{G}^\ast_e(\mathbf{r}_{A}, \mathbf{r}', \omega') \cdot
\hat{\mathbf{f}}^\dagger(\mathbf{r}', \omega', t') \right] \,.
\label{eq:nonlineqnofmotion}
\end{gather}
\end{widetext}
Equation~(\ref{eq:nonlineqnofmotion}) is now a nonlinear operator
equation that is capable of describing resonant processes involving
two polaritons. This is despite the fact that the original Hamiltonian
(\ref{eq:hamiltonian}) is bilinear in all operators. The effective
nonlinearity appears as a consequence of the iteration. In order to
pick out the resonant interactions from the equation of motion, we
introduce slowly varying amplitude operators as
$\hat{\tilde{\mathbf{f}}}(\mathbf{r},\omega, t)
=\hat{\mathbf{f}}(\mathbf{r}, \omega, t) e^{i \omega t}$ and
$\hat{\tilde{A}}_{mn}(t)=\hat{A}_{mn}(t) e^{-i \omega_{mn} t}$ and
apply the Markov approximation. This involves taking the slowly
varying amplitude operators out of the integral at the upper time $t$.
For simplicity let us demonstrate this for one of the terms in
Eq.~(\ref{eq:nonlineqnofmotion}),
\begin{gather}
\mathbf{I}_1(\mathbf{r}, \omega, t) \\
\equiv
\frac{1}{\hbar^2}
\sum_{k,m,n} \int\limits_0^t d t' \int\limits_0^\infty d\omega' \int d^3r' \,
e^{i \omega_{mn}(t-t')} \hat{A}_{kn}(t') \nonumber\\
\times \mathbf{d}_{mn} \cdot \bm{G}^\ast_e(\mathbf{r}_{A},\mathbf{r},\omega) 
\mathbf{d}_{km}
\cdot \bm{G}_e(\mathbf{r}_{A}, \mathbf{r}',\omega') \cdot
\hat{\mathbf{f}}(\mathbf{r}',\omega', t')\nonumber\\
=
\frac{1}{\hbar^2}
\sum_{k,m,n} \int\limits_0^\infty
d\omega' \int d^3r' \, 
\mathbf{d}_{mn} \cdot
\bm{G}^\ast_e(\mathbf{r}_{A}, \mathbf{r},\omega) \hat{A}_{kn}(t) \nonumber \\
\times  \mathbf{d}_{km} \cdot
\bm{G}_e(\mathbf{r}_{A}, \mathbf{r}',\omega') \cdot \hat{\mathbf{f}}
(\mathbf{r}',\omega', t) \int\limits_{0}^{t}d t' e^{i(\omega_{mk} +
\omega')(t-t')} \, . \nonumber
\end{gather}
The integrals can be approximated in the long-time limit, i.e. by
extending the upper limit of integration to infinity and assuming
that the atomic transitions are well away from the field resonances,
so that
$\int_{0}^{t} d t' e^{i(\omega_{mk}+\omega')(t-t')} \sim -\frac{1}{i
(\omega_{mk} + \omega')}$. This leads to the result
\begin{gather}
\mathbf{I}_1(\mathbf{r}, \omega, t)
=
\frac{1}{\hbar^2}
\sum_{k,m,n} \int\limits_0^\infty d\omega' \int d^3r' \, 
\mathbf{d}_{mn} \cdot
\bm{G}^\ast_e(\mathbf{r}_{A}, \mathbf{r},\omega)  \nonumber \\ 
\times  \frac{\hat{A}_{kn}(t) \mathbf{d}_{km} \cdot
\bm{G}_e(\mathbf{r}_{A}, \mathbf{r}',\omega')}{i (\omega_{mk} +
\omega')}\hat{\mathbf{f}} (\mathbf{r}',\omega',t) .
\end{gather}
The other three terms in Eq.~(\ref{eq:nonlineqnofmotion}) can be
approximated in an analogous way.

\begin{figure}
\includegraphics[width=5cm]{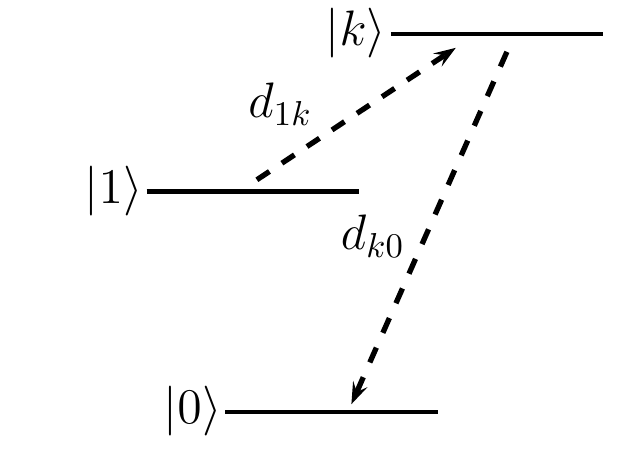}
 \caption{The transition from $\ket{1}$ to $\ket{0}$ is mediated by  the
virtual transitions via the state $\ket{k}$.}
 \label{trans_coupling}
\end{figure}

For our present investigation one has to keep in mind that in the 
nonretarded limit the polariton spectrum is not continuous (see 
discussion in Sec.~\ref{secIIIA}) but consists of of a quasidiscrete 
set of lines of midfrequencies
$\Omega_\nu$ and widths $\gamma_\nu$, where the linewidths are typically very
much smaller than the line center separations
$\gamma_\nu\ll(\Omega_{\nu+1}-\Omega_{\nu-1})/2$. We then divide the $\omega$
axis into intervals $\Delta_\nu = [(\Omega_{\nu-1}+\Omega_\nu)/2,
(\Omega_\nu+\Omega_{\nu+1})/2]$.
Recalling the resonance condition $\Omega_{1} \approx \omega_{10} + \Omega_2$,
we apply the rotating-wave approximation and finally arrive at the effective
equation of motion describing the dynamics of the resonant atom--polariton
coupling where now the frequency integrals have to be taken over the linewidth
of the surface polaritons,
\begin{gather}
\dot{\hat{\mathbf{f}}}(\mathbf{r}, \omega) = -i \omega
\hat{\mathbf{f}}(\mathbf{r}, \omega) \nonumber \\
 -i \int d^{3} r' \int\limits_{\Delta_\nu} d \omega' \, \left[
\hat{\mathbf{g}}(\mathbf{r}, \mathbf{r}', \omega,\omega') \cdot
\hat{\mathbf{f}}(\mathbf{r}', \omega') \right]\,.
\end{gather}
Here we have introduced the abbreviation
\begin{gather}
\hat{\mathbf{g}}(\mathbf{r}, \mathbf{r}', \omega,\omega') = \nonumber \\
-\frac{\hat{A}_{10}}{\hbar} \sum_{k} \bigg[ \frac{ \mathbf{d}_{k0} \cdot
\bm{G}^\ast_{e}(\mathbf{r}_{A}, \mathbf{r}, \omega) \otimes \mathbf{d}_{1k}
\cdot \bm{G}_{e}(\mathbf{r}_{A}, \mathbf{r}', \omega') }{\omega_{k1} + \omega'}
\nonumber \\ 
-\frac{ \mathbf{d}_{1k} \cdot \bm{G}^\ast_{e}(\mathbf{r}_{A},\mathbf{r},
\omega) \otimes \mathbf{d}_{k0} \cdot \bm{G}_{e}(\mathbf{r}_{A}, \mathbf{r}',
\omega') }{\omega_{0k} + \omega'} \bigg] \,
\end{gather}
for the operator-valued coupling tensor. As one can see from the structure of
$\hat{\mathbf{g}}(\mathbf{r},\mathbf{r'},\omega,\omega')$, the atom--polariton
coupling is mediated by a virtual atomic transition from $\ket{1}\rightarrow
\ket{0}$  via an intermediate state $\ket{k}$ (see Fig.~\ref{trans_coupling})
with dipole moments $\mathbf{d}_{1k}$ and $\mathbf{d}_{0k}$.
The equation of motion (\ref{eq:nonlineqnofmotion}) can be thought of as being
generated by the effective second order interaction Hamiltonian

\begin{equation}
 \dot{\hat{\mathbf{f}}} = (i \hbar)^{-1}  [\hat{\mathbf{f}}, \hat{H}_{F}
+\hat{H}_{\mathrm{(int)eff}}] \, ,
\end{equation}
where
\begin{gather}
\hat{H}_{\mathrm{(int)eff}} = \hbar \int d^{3} r' \int d^{3} r
\int\limits_{\Delta_\nu} d \omega \int\limits_{\Delta_{\nu'}} d \omega'
\, \nonumber \\ 
\hat{\mathbf{f}}^\dagger(\mathbf{r},\omega) \cdot
\bigl[\hat{\mathbf{g}}(\mathbf{r},\mathbf{r}', \omega,\omega')
-\hat{\mathbf{g}}^\dagger(\mathbf{r},\mathbf{r}',
\omega,\omega')\bigr]
 \cdot
\hat{\mathbf{f}}(\mathbf{r}', \omega')\,.
\label{eq:effectiveH}
\end{gather}
This Hamiltonian describes the effective creation of one polariton
excitation with a simultaneous annihilation of another. In the
specific scenario depicted in Fig.~\ref{ourpic}, only the term
involving $\hat{\mathbf{g}}(\mathbf{r},\mathbf{r}', \omega,\omega')$
will contribute to the near-resonant interaction Hamiltonian.

\subsection{Coupling to singly excited polaritons \label{secIIIA}}

We consider an atom at nonretarded distance $z$ from a flat surface
of multi--resonance Drude--Lorentz permittivity
\begin{equation}
\label{eq:Drude}
\varepsilon (\omega) = 1 + \sum_{j}
\frac{\omega_{Pj}^{2}}{\omega_{Tj}^{2} - \omega^{2} - i \omega \Gamma_{j}}
\end{equation}
with plasma frequencies $\omega_{Pj}$ and transverse resonance
frequencies $\omega_{Tj}$. 
The Green tensor for a half-infinite dielectric medium (subscript $d$)
in vacuum (subscript $v$) can be given as

\begin{gather}
 \mathbf{G} (\mathbf{r}, \mathbf{r}, \omega) = \frac{i}{8 \pi} \int_{0}^{\infty} d k_{\rho} \frac{k_{\rho}}{k_{vz}} e^{2 i k_{vz} z_{A}} \left\lbrace r_s (\omega)\begin{pmatrix}
  1 & 0 & 0 \\
  0 & 1 & 0 \\
  0 & 0 & 0
 \end{pmatrix} \right. \nonumber \\
 \left.
 + r_p (\omega) \frac{c^{2}}{\omega^{2}} \begin{pmatrix}
  -k_{vz}^{2} & 0 & 0 \\
  0 & -k_{vz}^{2} & 0 \\
  0 & 0 & 2 k_{\rho}^{2}
 \end{pmatrix} \right\rbrace
\end{gather}
where
\begin{gather}
r_p(\omega)=\frac{\varepsilon(\omega)k_{vz}-k_{dz}}{\varepsilon(\omega)k_{vz}+k_{dz}}, 
r_s(\omega)=\frac{k_{vz}-k_{dz}}{k_{vz}+k_{dz}}
\end{gather}
are the Fresnel reflection coefficients for $s-$ and $p-$polarized waves. 
In the nonretarded limit the approximation $k_{vz}=k_{dz}=ik_{\rho}$
can be made and the Green tensor of such a surface
reduces to
$\bm{G}(\mathbf{r},\mathbf{r},\omega)\simeq
z^{-3}\bm{G}'(\omega)$ with
\begin{equation}
\label{eq:Gprime}
\bm{G}'(\omega) = \frac{c^{2}}{32 \pi \omega^{2}} \tilde{r}_p(\omega)
 \begin{pmatrix}
  1 & 0 & 0 \\
  0 & 1 & 0 \\
  0 & 0 & 2
 \end{pmatrix}
\end{equation}
where now $\tilde{r}_p(\omega)=(\varepsilon(\omega)-1)/(\varepsilon(\omega) +1)$.
The general condition to obtain $p-$polarized surface waves is given
by the dispersion relation 
\begin{equation}
\varepsilon(\omega)k_{vz}+k_{dz}=0 \,.
\end{equation}
When taken to the nonretarded limit ($k \to \infty$) it exhibits resonances 
where the associated modes are the surface polaritons (strictly speaking, 
there are poles in the complex frequency plane where $\varepsilon(\omega)=-1$).

Combining these two equations, one sees that the local density of
states $\omega^2\mathrm{Im}\,\bm{G}(\mathbf{r},\mathbf{r},\omega)$
near a given polariton resonance can be approximated by a single
Lorentzian peak of mid-frequency $\Omega_\nu$ and width $\gamma_\nu$,
\begin{equation}
\omega^2 \mathrm{Im}\,\bm{G}(\mathbf{r}, \mathbf{r}, \omega) \simeq
\Omega_\nu^2 \mathrm{Im}\,\bm{G}(\mathbf{r}, \mathbf{r}, \Omega_\nu)
\frac{\gamma_\nu^2/4}{ (\omega-\Omega_\nu)^2+\gamma_\nu^2/4 }\,.
\label{Lorentzshape}
\end{equation}
It is then useful to define the respective single--polariton
excitations (similar to the generic construction of
quantum-mechanical single-photon wave
packets, cf. Ref.~\cite{PRA77_012110_2008}) as
\begin{gather}
\ket{1(\mathbf{r}_{A}, \Omega_{i})} = \sqrt{\frac{2}{\pi \gamma_{i}}} 
\int\limits_{\Delta_{\Omega_{i}}} d \omega \int d^3s\,
\frac{1}{g(\mathbf{r}_{A}, \Omega_{i})}  \nonumber \\
 \times \bm{G}^\ast_{e}(\mathbf{r}_{A}, \mathbf{s}, \omega) \cdot
\hat{\mathbf{f}}^{\dagger}(\mathbf{s}, \omega) \ket{\left\{ 0\right\} }
\label{statenewdef}
\end{gather}
with the normalization factor
\begin{equation}
g(\mathbf{r}_{A}, \Omega_{i}) = \sqrt{\frac{\mu_{0}}{\hbar \pi} \Omega_{i}^{2}
\mathrm{Tr}\, \mathrm{Im}\,\bm{G}(\mathbf{r}_{A},\mathbf{r}_{A},
\Omega_{i}) }\,, 
\end{equation}
where $\mathrm{Tr}$ denotes the trace.
Using the integral relation (\ref{eq:magicformula}) for the Green
tensor,  the integral in frequency can be approximated in the
long--frequency limit by extending the upper limit of integration to
infinity using the definition for normalization of a Lorentzian
function
\begin{equation}
\int_{-\infty}^{\infty} \frac{1}{\pi}
\frac{\gamma/2}{(\omega-\Omega)^{2} + \gamma^{2}/4}=1,
\end{equation}
one easily checks that the states (\ref{statenewdef}) are indeed
properly normalized, $\langle 1(\mathbf{r}_{A},
\Omega_{i})\ket{1(\mathbf{r}_{A},\Omega_{i})}=1$. Note that the states
$\ket{1(\mathbf{r}_{A}, \Omega_{i})}$ carry a vector index as well as
the continuous space and frequency labels.

In our envisaged situation of a resonant coupling between a single atomic
transition and the difference between two polariton resonances, the energies of
the initial and final states are identical. Degenerate first-order perturbation
theory asserts that the interacting potential is \cite{CohenBookQuantum}
\begin{equation}
U_{\mathrm{eff}} = \sqrt{ \md{\bra{K} \hat{H}_{\mathrm{eff}} \ket{I}}^{2}}.
\label{Ueff}
\end{equation}
Here $\ket{I}=\ket{1_{A}} \ket{0_{1}} \ket{1_{2}}$ stands for the tensor
product of the initial excited atomic state $\ket{1_{A}}$ and a singly excited
polariton with frequency $\Omega_{2}$, and
$\ket{K}=\ket{0_{A}}\ket{1_{1}}\ket{0_{2}}$ denotes the tensor product of the
final atomic state $\ket{0_{A}}$ and a single excitation in the polariton with
frequency $\Omega_{1}$. The single-polariton states
$\ket{1_{\nu}}\equiv\ket{1(\mathbf{r}_A,\Omega_\nu)}$ are defined according to
Eq.~\eqref{statenewdef} and  $\ket{0_{\nu}}$ denotes the polariton ground state
$\ket{0_{\nu}} = \ket{\{0\}}$, $\forall\omega\in
[\Omega_{\nu}-\delta\omega/2,\Omega_{\nu}+\delta\omega/2]$.

Using the commutation rules of the operators as well as the properties of the
Green functions, together with the definition of the Lorentzian lineshape, Eq.
\eqref{Lorentzshape}, we find that the effective interaction potential can be
written in the form
\begin{gather}
U_{\mathrm{eff}} = - \frac{\mu_{0} \Omega_{1} \Omega_{2}}{2} \nonumber \\
\times \sqrt{\frac{\gamma_{1} \gamma_{2}}{\mathrm{Tr} \left[ \mathrm{Im}
\bm{G}(\mathbf{r}_A,\mathbf{r}_A,\Omega_1)\right] \mathrm{Tr} \left[ \mathrm{Im}
\bm{G}(\mathbf{r}_A,\mathbf{r}_A,\Omega_2) \right] }} \nonumber \\
\sum_{k} \Bigg\{ \mathrm{Tr} \left[  \mathrm{Im}
\bm{G}(\mathbf{r}_A,\mathbf{r}_A,\Omega_1) \cdot \mathbf{d}_{0k} \otimes
\mathbf{d}_{k1} \cdot \mathrm{Im} \bm{G}(\mathbf{r}_{A},\mathbf{r}_{A},\Omega_2)
\right]  \nonumber \\ \times
\frac{\Omega_{1}+\omega_{0k}}{(\Omega_{1}+\omega_{0k})^2+\gamma_1^2/4} \nonumber \\
- \mathrm{Tr} \left[ \mathrm{Im}\bm{G}(\mathbf{r}_A,\mathbf{r}_A,\Omega_1)
\cdot \mathbf{d}_{k1} \otimes \mathbf{d}_{0k} \cdot \mathrm{Im}
\bm{G}(\mathbf{r}_A,\mathbf{r}_A,\Omega_2) \right] \nonumber \\ \times
\frac{\Omega_{1} + \omega_{k1}}{(\Omega_{1} + \omega_{k1})^2 + \gamma_1^2/4} \Bigg\} .
\label{eq20}
\end{gather}
Let us compare Eq.~\eqref{eq20} with the nonresonant Casimir--Polder
potential at finite temperature \cite{acta2008,AlexPRA2010},
\begin{gather}
\Delta E^{\mathrm{NR}}  =  \mu_{0} k_{B} T \sideset{}{'}\sum_{j=0}^{\infty}
\xi_{j}^{2} \mathrm{Tr} \left[ \bm{\alpha}(i\xi_{j}) \cdot \bm{G}
(\mathbf{r}_{A},\mathbf{r}_{A},i \xi_{j}) \right] \nonumber \\
+ \mu_{0} \sum_{k \neq n} \omega_{kn}^{2}  \bar{n}_{\mathrm{th}} (\omega_{kn})
\mathbf{d}_{nk} \cdot \mathrm{Re} \,\bm{G}
(\mathbf{r}_{A},\mathbf{r}_{A},\omega_{kn}) \cdot \mathbf{d}_{kn},
\label{U_CP}
\end{gather}
where $\xi_{j}$ are the Matsubara frequencies, $\bm{\alpha}(\omega)$ is
the atomic polarizability and
$\bar{n}_{\mathrm{th}}(\omega)=[\exp(\hbar\omega/k_BT)-1]^{-1}$ the
thermal occupation number. We note that the effective potential
scales with the atom-surface distance $z$ in exactly the same way as
nonresonant Casimir--Polder potential ($\propto z^{-3}$).
The effective Hamiltonian \eqref{eq:effectiveH} is quadratic in the
field variables and contributes to the potential \eqref{Ueff} at  
(degenerate) first-order perturbation theory. The nonresonant
potential arises from a Hamiltonian which is linear in the field
variables, contributing only in second-order perturbation theory
\cite{acta2008}. In both cases, we therefore obtain a result that is
quadratic in the atom-field coupling, or, equivalently, linear in the
imaginary part of the Green tensor (the local mode density).

The total potential experienced by the atom is the sum  of the nonresonant
(attractive) Casimir--Polder potential and the resonant coupling between the
atoms and the surface polaritons,
\begin{equation}
\Delta E^{\mathrm{Total}} = \Delta E^{\mathrm{NR}}+\Delta E^{\mathrm{R}}.
\end{equation}
Using Eqs.~(\ref{eq:Gprime}) and (\ref{Lorentzshape}),
the respective energy shifts for the nonresonant and (second-order) resonant
interactions in the nonretarded limit considering an isotropic atom are
\begin{gather}
\Delta E^{\mathrm{NR}}  =  -\frac{\mu_{0} c^{2} k_{B} T}{12 \pi \hbar z^3}
\sum_{k} \md{\mathbf{d}_{nk}}^{2}   \sideset{}{'}\sum_{j=0}^{\infty}
\frac{\omega_{kn}}{\omega_{kn}^{2} + \xi_{j}^{2}} \frac{\varepsilon (i \xi_{j})
-1}{\varepsilon (i \xi_{j}) +1} \nonumber \\
+ \frac{\mu_{0} c^{2}}{24 \pi z^3} \sum_{k} \bar{n}_{\mathrm{th}} (\omega_{kn})
\md{\mathbf{d}_{nk}}^{2} \mathrm{Re} \left[ \frac{\varepsilon (\omega_{kn})
-1}{\varepsilon (\omega_{kn}) +1} \right]
\label{CPeq}
\end{gather}
and
\begin{gather}
 \Delta E^{\mathrm{R}} = - \frac{\mu_{0} \Omega_{1} \Omega_{2}}{2 z^3}  
\sqrt{\frac{\gamma_{1} \gamma_{2}}{\mathrm{Tr} \left[ \mathrm{Im}   \bm{G}'
(\Omega_1) \right] \mathrm{Tr} \left[  \mathrm{Im} \bm{G}' (\Omega_2) \right]} }
 \nonumber \\ 
\times  \sum_{k} \Bigg\{ \mathrm{Tr} \left[ \mathrm{Im}  \bm{G}'(\Omega_1)
\cdot \mathbf{d}_{0k} \otimes  \mathbf{d}_{k1} \cdot \mathrm{Im}
\bm{G}'(\Omega_2) \right]  \nonumber \\
\times  \frac{\Omega_{1}+\omega_{0k}}{(\Omega_{1}+\omega_{0k})^2+\gamma_1^2/4}  \nonumber \\
 - \mathrm{Tr} \left[ \mathrm{Im}\bm{G}' (\Omega_1) \cdot  \mathbf{d}_{k1}
\otimes \mathbf{d}_{0k} \cdot   \mathrm{Im} \bm{G}'(\Omega_2) \right]  \nonumber
\\
 \times  \frac{\Omega_{1} + \omega_{k1}}{(\Omega_{1} + \omega_{k1})^2
+  \gamma_1^2/4} \Bigg\},
\label{CPres}
\end{gather}
recall Eq.~(\ref{eq:Gprime}).

\subsection{Thermal States}
 
As we are dealing with thermally excited surface polaritons the concept of
perturbation theory has to be expanded from pure states described by a single
state vector to a statistical mixture or ensemble of states \cite{RMP29_1_1957}.
The density matrix for a thermal state with temperature $T$ can be written in
the Fock basis $|n\rangle$ as
\begin{equation}
\hat{\rho}_{\mathrm{th}} = \sum_{n} p_{n} \ket{n} \bra{n} = \sum_{n}
\frac{e^{-n \hbar \Omega_{n} / k_{B} T}}{Z(T)} \ket{n} \bra{n}
\end{equation}
where $Z(T)=\sum_m e^{-m \hbar\Omega_{m}/k_{B}T}$ denotes the partition function.

So far we have computed the interaction energy for the situation in which there
is initially only one excited polariton with frequency $\Omega_{2}$ and in the
final state only one polariton with frequency $\Omega_{1}$ [see
Eq.~\eqref{Ueff}]. This has to be generalized to thermal states in which there
can be initially $m$ polaritons with frequency $\Omega_{1}$ and $n$ polaritons
with frequency $\Omega_{2}$. In this case, we rewrite the result of the
perturbation theory as 
\begin{equation}
\md{\bra{K} \hat{H}_{\mathrm{eff}} \ket{I}}^{2} = \mathrm{Tr}
\left[\hat{H}_{\mathrm{eff}} \hat{\rho}_{\mathrm{in}} \hat{H}_{\mathrm{eff}}
\ket{K} \bra{K} \right]
\end{equation}
where
\begin{gather}
\hat{\rho}_{\mathrm{in}} = \hat{\rho}_{\mathrm{th}}(\Omega_1) \otimes
\hat{\rho}_{\mathrm{th}}(\Omega_2) \otimes \ket{1_{A}} \bra{1_{A}} \nonumber \\
= \sum_{n,m} p_{m}^{(1)} p_{n}^{(2)}  \ket{m_{1}, n_{2}}  \bra{m_{1}, n_{2}}
\otimes \ket{1_{A}} \bra{1_{A}} .
\end{gather}
Due to the form of the effective interaction Hamiltonian
$\hat{H}_{\mathrm{(int)eff}} \propto \hat{\mathbf{f}}^\dagger(\mathbf{r},\omega)
\hat{\mathbf{f}}(\mathbf{r}', \omega') $ the only final state $\ket{K}$ that
provides a non-vanishing matrix element will be $ \ket{K} = \ket{(m+1)_{1},
(n-1)_{2}} \ket{0_{A}}$. Recall that $\hat{\mathbf{f}} \ket{k} = \sqrt{k}
\ket{k-1}$ and $\hat{\mathbf{f}}^{\dagger} \ket{k} = \sqrt{k+1} \ket{k+1}$.
Hence,
\begin{eqnarray}
\md{\bra{K} \hat{H}_{\mathrm{eff}} \ket{I}}^{2} &=& \sum_{m,n}  p_{m}^{(1)}
p_{n}^{(2)} (m+1)_{(1)} (n)_{(2)} U_{\mathrm{eff}}^{2} \nonumber \\
 &=& \left[ \bar{n}_{\mathrm{th}}(\Omega_{1}) +1 \right]
\bar{n}_{\mathrm{th}}(\Omega_{2}) U_{\mathrm{eff}}^{2}.
\end{eqnarray}
Finally, the resonant energy shift $\Delta E^{R}$ for thermal states will be given as
\begin{equation}
\label{Thermal}
 \Delta E^{\mathrm{R}} = U_{\mathrm{eff}} \sqrt{ \left[
\bar{n}_{\mathrm{th}}(\Omega_1) +1 \right] \bar{n}_{\mathrm{th}}(\Omega_2) }.
\end{equation}
This result is intuitively clear, as the initial polariton with frequency
$\Omega_2$ has to be thermally populated before the resonant interaction can
take place.

\subsection{Discussion}

Let us apply the general results for the potentials (\ref{CPeq}) and
(\ref{Thermal}) with (\ref{CPres}) to the envisaged Drude--Lorentz model 
(\ref{eq:Drude}). For this scenario with two well-separated narrow polariton
resonances, 
(in this case the width of the polariton resonance $\gamma$ is approximately the
same as the width of the Drude-Lorentz resonance $\Gamma$ of the material), the
effective potential becomes
\begin{gather}
\Delta E^{\mathrm{R}}  \sim - \frac{\mu_{0} c^{2}}{128 \pi z^{3}}
\frac{\omega_{P1} \omega_{P2}}{ \sqrt{\Omega_{1} \Omega_{2}} }
\nonumber \\
\times \sqrt{ \left[ \bar{n}_{\mathrm{th}}(\Omega_1) +1 \right]
\bar{n}_{\mathrm{th}}(\Omega_2) } \sum_{k} \frac{5
\md{\mathbf{d}_{0k}} \md{\mathbf{d}_{k1}} }{12} \nonumber \\
\times \Bigg\{
\frac{\Omega_{1}+\omega_{0k}}{(\Omega_{1}+\omega_{0k})^2+\gamma_1^2/4}
- \frac{\Omega_{1} + \omega_{k1}}{(\Omega_{1} + \omega_{k1})^2 +
\gamma_1^2/4} \Bigg\} .
\label{eq32}
\end{gather}

Similarly, the nonresonant Casimir--Polder potential for a
one-polariton model is
\begin{gather}
\Delta E^{\mathrm{NR}}  \sim - \frac{\mu_{0} c^{2}}{48 \pi z^{3}}
\frac{k_{B} T}{ \hbar} \sum_{\nu} \frac{ \omega_{P}^{2}
\md{\mathbf{d}_{1 \nu}}^{2}}{\Omega \omega_{\nu 1}}\nonumber \\
+\frac{\mu_{0} c^{2}}{ 24 \pi z^{3}} \sum\limits_\nu
\bar{n}_{\mathrm{th}} (\omega_{\nu 1}) \md{\mathbf{d}_{1 \nu}}^{2}
\nonumber \\
\times  \mathrm{Re} \left[ \frac{\omega_{P}^{2}}{2(\omega_{T}^{2} -
\omega_{1 \nu}^{2} -i \omega_{1 \nu} \Gamma) + \omega_{P}^{2}}
\right].
\end{gather}
For typical cell materials such as sapphire \cite{sapphire} and quartz
\cite{quartz} the resonant second order shift was evaluated numerically
for atoms typically used in these type of experiments such as Rubidium.
For the temperatures at which these experiments are performed, from
350--600~K, the surface polaritons frequencies are thermally populated.
In comparison to the nonresonant CP shift (which is in the order of 
several GHz for Rydberg atoms) the resonant second order shift is too 
small (only several kHz) to be relevant. Although we have detailed 
experimental results in Ref.~\cite{Kubler2010}, a comparison between
theory and the experimental work cannot be performed because 
we lack information on the real cell material properties.

Let us investigate which intermediate atomic transitions might provide
the largest second order effect. In order to have an effect that is
comparable to then nonretarded Casimir--Polder interaction, there has
to be a matching atomic transition between energetically close states
--- note that the transition $\ket{1} \to \ket{0}$ does not need to be
allowed by the selection rules --- i.e. the intermediate state
$\ket{k}$ has to be close to the initial and final states $\ket{1}$
and $\ket{0}$, see Fig.~\ref{trans_coupling}. The reason for this
constraint is the rapidly decreasing magnitude of the dipole
transition matrix element between states with increasing energy
difference. Therefore, the dominant contribution will come from an
intermediate state $\ket{k}$ approximately halfway between the initial and
final states.

In this case the difference between the resonant and nonresonant terms
will come from the last line in Eq.~\eqref{eq32}. Its maximum value is
obtained whenever $\omega_{0k}$ or $\omega_{1k}$ is one of $\Omega_{1}
\pm \gamma_{1}/2$; away from these points the numerical value of this
term decreases. As we have assumed throughout our calculations that
all atomic transitions are far from any single-polariton resonance,
the Lorentzian peaks have to be broad, i.e. $\gamma_1$ has to be
large. This in turn means that, in order for this second order effect to
be comparable to the nonresonant Casimir--Polder potential, a strongly
dissipative material is needed. Note that we assumed in our derivation
that linewidths need to smaller than the line center separations which
does not exclude the possibility of the peaks to be broad, in fact
that is a characteristic that one observes in real polariton spectra.

With these considerations in mind, we give some estimates to show that it would
possible to access this phenomenon. For our purpose we choose the $27S_{1/2}\to
26S_{1/2}$ transition in rubidium. In order for the second-order process to be
relevant, one has to find a material with surface polariton frequencies whose
difference matches that atomic transition ($\sim 17$~cm$^{-1}$). For example,
let us choose a material with surface polaritons at 73 and 90 cm$^{-1}$ (which
we model as two narrow resonances with $\gamma \sim 0.03 \Omega$, in which case
$\mathrm{Im} r_{p} (\Omega) > 100$). With an atom-surface distance of $z=1
\mu$m, the Casimir--Polder shift due to the second-order process is $\Delta
E^{\mathrm{R}} =-2.74619 \times 10^7 s^{-1}$. As the total level shift can be
calculated to be $\Delta E^{\mathrm{Total}}=-1.07583\times 10^8 s^{-1}$, the
resonant second-order process contributes around $25\%$ to the todal
Casimir--Polder shift and is thus expected to be experimentally accessible.

\section{Conclusion}
\label{sec:conclusions}

We have shown that second order effective interactions involving two
surface polaritons lead to novel contributions to dispersion
interactions such as the Casimir-Polder potential. We have explicitly
derived the dependence of such effects on the atomic and surface
parameters and compared their magnitude to that of the conventional
nonresonant Casimir-Polder interaction. In principle, one can
envisage conditions under which the two become comparable.

However, a more quantitative analysis is largely dependent on exact
data of the individual surface properties. For each dielectric
material there is a unique surface polariton spectrum that depends
sensitively on the concentration and distribution of the impurities
and surface quality of the samples (surface roughness). As each sample
is unique, the polariton spectrum should be found
experimentally. Current experimental findings have revealed some
discrepancies, which are most likely due to variations in the quality
of the sample, the degree of its impurities and the orientation of the
crystal axes \cite{EPJD15_2001}. The latter effect provides a handle
to tune the surface polariton frequency by changing the crystal
orientation \cite{PRL83_26_1999}.

\acknowledgments

We would like to  acknowledge fruitful discussions with C.S.~Adams,
H.~K\"ubler and T.~Pfau. SR is supported by the PhD grant
SFRH/BD/62377/2009 from FCT, co-financed by FSE, POPH/QREN and EU.
This work was partially supported by the UK Engineering and
Physical Sciences Research Council.

\bibliography{article_SRibeiro}
\end{document}